\documentclass[pre,floatfix,superscriptaddress,twocolumn,showpacs]{revtex4-1}
\usepackage{graphicx}
\usepackage{amsmath,amssymb,amsfonts}
\usepackage{mathtools}
\usepackage{bm}
\usepackage{color}
\usepackage[colorinlistoftodos]{todonotes}

\begin{document}
\title{Rejuvenation and memory effects in active glasses induced by thermal and active cycling}
\author{Giulia Janzen}
\author{Liesbeth M.~C.~Janssen}
\email{l.m.c.janssen@tue.nl}
\affiliation{Department of Applied Physics, Eindhoven University of Technology, P.O.~Box 513, 5600 MB Eindhoven, The Netherlands}

\date\today

\begin{abstract}
It has recently been shown that thermal active glasses can display physical aging behavior comparable to that of passive glasses, although there are some notable distinctions due to the intrinsic non-equilibrium nature of active matter. The question whether active disordered materials can also exhibit rejuvenation and memory effects, akin to the phenomenology of e.g.\ spin glasses, has thus far remained unexplored. Here we address this question by numerical simulations of active glasses composed of active Brownian particles that are subjected to a thermal or active cycling protocol. We find that an active system undergoing thermal cycling indeed shows rejuvenation and memory effects, with the strength of rejuvenation depending on the persistence time. In contrast, however, a passive Brownian system subjected to the same thermal cycle lacks the rejuvenation effect. We attribute this to the enhanced motility of active particles, which enables them to escape from their cages and restart aging at the new temperature, thus rejuvenating the material. Finally, we also demonstrate that both rejuvenation and memory effects can be induced by an activity cycle which quenches the material from an active to passive glass and back, providing a unique means to rejuvenate active matter. 
%It has been recently established that dense active matter shares similarities with conventional glassy phenomena. Moreover, thermal active glasses display an aging behavior that is comparable to passive systems, although there are some distinctions. In spin glasses, a system subjected to a thermal cycle shows rejuvenation and memory effects. However, it is unknown how such a protocol would impact the dynamics of an active system. Here we show, by means of numerical simulations, that an active system subjected to a thermal cycle exhibits rejuvenation and memory effects, with the strength of rejuvenation depending on persistence time. In contrast, a passive system subjected to the same thermal cycle lacks the rejuvenation effect, which may result from a small temperature jump or brief protocol duration. In the active case, instead, activity enables particles to escape from their cages and restart aging at the new temperature, leading to rejuvenation. Furthermore, we perform an active cycle from an active to a passive glass, showing that rejuvenation and memory are present.
\end{abstract}
\maketitle
\section{Introduction}
The dynamics of many densely disordered systems, i.e.\ glassy materials, is characterized by an extremely slow structural relaxation that is often explicitly dependent on the age of the material. 
%The dynamics of many disordered systems, e.g. glasses, is time-dependent and is usually characterized by an extremely slow relaxation. 
This age (or waiting time) dependence is called physical aging and is particularly well studied for glasses following a sudden quench toward a lower temperature~\cite{binder2011glassy,biroli2013perspective,debenedetti2001supercooled,PhysRevX.7.031028}. After such a temperature quench, the structural relaxation time tends to increase with the age of the material, typically as a power law \cite{kob1997aging,foffi2004aging,hodge1995physical,berthier2009statistical,lunkenheimer2005glassy,zhao2013using,Raty2015,Wang2006,odegard2011physical,martin1993aging,mckenna1995evolution}. Theoretically, this aging behavior can be understood as a gradual approach of the material toward lower-energy equilibrium states \cite{debenedetti2001supercooled}. While a single thermal quench is one of the most popular protocols to study the aging behavior of glassy systems, other protocols, such as thermal cycling involving repeated temperature changes, may lead to even richer dynamical behaviors and offer more versatility to characterize a material's out-of-equilibrium dynamics. 

%One of the most popular protocol to study the aging dynamics consists of a temperature quench toward a lower temperature \cite{kob1997aging,foffi2004aging}. After such a temperature quench, the structural relaxation time increases with the age of the system \cite{hodge1995physical,berthier2009statistical,lunkenheimer2005glassy,zhao2013using,Raty2015,Wang2006,odegard2011physical,martin1993aging,mckenna1995evolution}. Although this aging protocol allows us to characterize the aging behavior of glassy systems, other experimental protocols, e.g. thermal cycling, could be used to better characterize their out-of-equilibrium dynamics. 

For spin glasses, the effect of thermal cycling has been studied theoretically, experimentally, and in numerical simulations, disclosing new insights into their dynamical behavior \cite{PhysRevB.38.373,koper1988domain,PhysRevLett.58.57,mezard1987spin,Jin2017,PhysRevLett.81.3243,PhysRevLett.51.911,PhysRevB.48.13977,PhysRevB.64.174204,PhysRevLett.89.097201,PhysRevB.66.054404,PhysRevB.69.184423,PhysRevB.71.214429,PhysRevB.65.024439,PhysRevB.72.064204,PhysRevLett.81.3243,Jonason2000,Sasaki2002,Baity-Jesi2023}. 
Briefly, the protocol consists of three steps: firstly, a high-temperature liquid is cooled to $T_{q_{1}}<T_g$ (where $T_g$ is the glass transition temperature); after a certain interval, the temperature is quenched once more to a lower temperature $T_{q_{2}}<T_{q_{1}}$; and finally, the temperature is reheated to $T_{q_{1}}$.
During the first step, the system undergoes aging. In the subsequent step, instead of the relaxation process slowing down further, the aging process restarts at the new temperature $T_{q_{2}}$. This phenomenon is commonly referred to as 'rejuvenation'. The relaxation at $T_{q_{2}}$ will be identical to that obtained from a direct quench at this temperature if 
%$\Delta T$ 
the temperature jump 
is sufficiently large \cite{PhysRevLett.81.3243,PhysRevB.64.174204,PhysRevB.65.024439,lefloch1992can,sasaki2002deviations}. In the third step of the thermal cycling protocol, when the temperature is brought back to $T_{q_{1}}$, the system exhibits \textit{full memory} if the relaxation restarts exactly from the point reached before the second step. However, achieving full memory requires a sufficiently large temperature difference, denoted as $\Delta T$. If $\Delta T$ is not large enough, the behavior observed in the third step will be influenced by the aging that occurred at $T_{q_{2}}$ \cite{vincent1995contrasting,Vincent2007}. 

%During the first step, the system undergoes aging. After the second temperature jump, aging restarts, resulting in rejuvenation. In the last step, when the temperature is increased, the system may remember the state it reached before the second temperature jump, exhibiting a memory effect.

%The rejuvenation and memory effects observed in spin glasses can be explained by a characteristic length that slowly increases with time following a temperature quench \cite{PhysRevB.38.373,koper1988domain,PhysRevLett.58.57}, or by using a hierarchical energy-landscape picture \cite{mezard1987spin,Parisi1980,PhysRevLett.35.1792,PhysRevLett.63.2853,Jin2017}. The occurrence of rejuvenation and memory effects relies on a clear separation of length and time scales \cite{PhysRevB.65.024439,PhysRevB.66.054404}.  According to the hierarchical picture, there is a continuously ramifying organization of metastable states defined as local minima of the coarse-grained free energy, each separated by finite barriers. The system is trapped in these local minima for a specific escape time that is related to the heights of the barriers. As the temperature drops, each minimum divides into new ones between which the system seeks a new equilibrium state (rejuvenation effect) \cite{vincent1995contrasting,PhysRevLett.81.3243} and the trapping time increases as the temperature decreases. When the temperature rises again, the new minima merge back into their ancestors (memory effect).

%Structural glass formers
Several studies have explored whether rejuvenation and memory, as observed in spin glasses, can also occur in structural glasses. Recent computer simulations of a continuously polydisperse model glassformer have shown that temperature cycling indeed leads to rejuvenation and memory effects, provided that both the duration of each step in the cycle and the temperature jumps between the steps are sufficiently large \cite{PhysRevLett.122.255502}. 
Similarly, in numerical simulations of binary Lennard-Jones mixtures, it was found that oscillatory temperature variations may induce rejuvenation, depending on the cooling rate and cycling amplitude \cite{PRIEZJEV2019131,PhysRevResearch.3.013234}.
Numerous studies have also explored emergent non-equilibrium phenomena in metallic glasses exposed to oscillatory temperature variations  \cite{SHANG2021116952,KETKAEW2020100,LIU201993}. Most notably, subjecting metallic glasses to cryogenic thermal cycling can induce rejuvenation and improve the plasticity of the material \cite{Ketov2015,GUO2018141,TONG2015240}.

The study of glassy phenomena has recently seen a renewed surge of interest through the advent of active matter, i.e., non-equilibrium systems composed of self-propelling particles.
Both theory and simulations \cite{szamel2016theory,feng2017mode,liluashvili2017mode,henkes2011active,ni2013pushing,berthier2013non,berthier2014nonequilibrium,szamel2015glassy,flenner2016nonequilibrium,berthier2017active,PhysRevX.6.011037,janssen2019active,berthier2019glassy}, as well as experiments \cite{Lozano2019,klongvessa2019active,klongvessa2019nonmonotonic,Bi2015,needleman2017active,parry2014bacterial,garcia2015physics,bi2016motility,nishizawa2017universal,angelini2011glass,Zhang2021}, have shown that dense active matter shares many similarities with conventional glassy systems, including anomalously slow dynamics and aging.
Thus far, the physical aging behavior of active systems has been explored in a small number of simulation studies for  thermal \cite{janzen2021aging} and athermal active glasses \cite{janssen2017aging,mandal2020multiple} using a single temperature or activity quench, respectively. Notably, for active thermal glasses, the aging relaxation dynamics was found to be governed by a time-dependent competition between thermal and active effects, with the active particles' persistence time controlling both the time scale and magnitude of the activity-enhanced speedup in dynamics \cite{janzen2021aging}. However, it remains unclear how a protocol such as a thermal cycle would impact the dynamics of an active glass, and whether activity itself could be used to design a new non-equilibrium protocol, such as an activity cycle.

Here we investigate how cyclic protocols affect the dynamics of structural active glasses. 
We find that, in contrast to our passive reference sample, active glasses do exhibit rejuvenation under a simple temperature cycle. Importantly, the strength of this thermally induced rejuvenation effect depends on the active particles' persistence time. % Specifically, in agreement with the fact that the long-time regime is dominated by activity, we observe that the strength of the rejuvenation effect in an active glass depends on the persistence time. 
We also observe a clear memory effect that, unlike rejuvenation, is more easily observed and independent of the persistence time.
%which in the active case  Our results show that, in contrast to rejuvention, the memory effect is also observed in the passive case. Furthermore, in the active case, we observed that this effect remains independent of the persistence time, the same memory effect is observed for both relatively small and relatively large persistence times. 
Moreover, we introduce a cyclic protocol unique to active matter, in which an active sample is made passive and then active again. By applying this protocol, we find that the rich out-of-equilibrium dynamics observed in the temperature cycle protocol becomes even more pronounced in the activity cycle. 

\section{Methods}
\subsection{Simulation model}
We study a two-dimensional (2D) binary mixture of thermal active Brownian particles (ABPs). The overdamped equations of motion for each particle $i$ are given by
\begin{align}
   &\gamma \,\dot{\bm{r}}_i=\sum_{i\ne j=1}^{N} \bm{f}_{ij}+f \, \bm{n}_i+\sqrt{2D_T} \, \bm{\eta} \label{1}\\ 
 &\dot{\theta}_i=\sqrt{2D_{r}} \,\eta_{\theta} \label{2}
\end{align}
where $\bm{r}_i=(x_i,y_i)$ and  $\theta_i$ represent the particle's spatial and rotational coordinates, respectively. The dots denote a time derivative. Thermal noise is modeled as an independent Gaussian stochastic process, $\bm{\eta}=(\eta_x,\eta_y)$, with zero mean and variance $2 D_T\delta(t-t^\prime)$, where $D_T=k_B T/\gamma$ with $k_B$ being the Boltzmann constant, $T$ the temperature, and $\gamma$ the friction coefficient. The rotational noise $\eta_{\theta}$ is a Gaussian stochastic process with zero mean and variance $2 D_r \delta(t-t^\prime)$. 
%The independent Gaussian stochastic processes $\bm{\eta}=(\eta_x,\eta_y)$ represents the thermal noise and has zero mean and variance $2k_BT/\gamma \delta(t-t^\prime)$, where $k_B$ is the Boltzmann constant, $T$ the temperature, and $\gamma$ a friction coefficient. In Eq.\ \ref{2}, $\eta_{\theta}$ is a Gaussian stochastic process with zero mean and variance $2 D_r \delta(t-t^\prime)$. 
The factors $D_T$ and $D_{r}$ represent the translational and rotational diffusion constants, respectively, and the ABP persistence time is defined as $\tau_r=D_r^{-1}$. The constant self-propulsion speed $f /  \gamma$ is applied to each ABP along a direction $\bm{n}_i=(\cos{\theta}_i,\sin{\theta}_i)$. Note that when the active force $f$ is equal to zero, Eq.\ \ref{1} reduces to the equation of motion for a passive Brownian particle. Finally, $\bm{f}_{ij}=-\nabla_i V(r_{ij})$ is the interaction force between particles $i$ and $j$, where $r_{ij}=\left|\textbf{r}_i-\textbf{r}_j \right|$ and $V$ is the Lennard-Jones potential with a cutoff distance $r_{ij} = 2.5 \sigma_{ij}$ and zero otherwise. In order to prevent crystallization we use the parameters of the 2D binary Kob-Andersen mixture \cite{kob1995testing}: $A=65 \%$, $B=35 \%$,  $\epsilon_{AA}=1$, $\epsilon_{BB}=0.5 \epsilon_{AA} $, $\epsilon_{AB}=1.5 \epsilon_{AA}$, $\sigma_{AA}=1$, $\sigma_{BB}=0.88 \sigma_{AA}$ and $\sigma_{AB}=0.8 \sigma_{AA}$. 
We set the density to $\rho=1.2$, the number of particles to $N=10\,000$ and $D_T=\gamma=1$. Results are in reduced units, where $\sigma_{AA}$, $\epsilon_{AA}$, $\frac{\sigma_{AA}^2 \gamma}{\epsilon_{AA}}$, and $\frac{\epsilon_{AA}}{k_B}$ are  the units of length, energy, time, and temperature, respectively. Simulations were performed using LAMMPS \cite{PLIMPTON19951} by solving Eqs.\ \ref{1} and \ref{2} via the Euler-Maruyama method with a step size $\delta t=10^{-4}$.
\begin{figure}
    \includegraphics [width=\columnwidth] {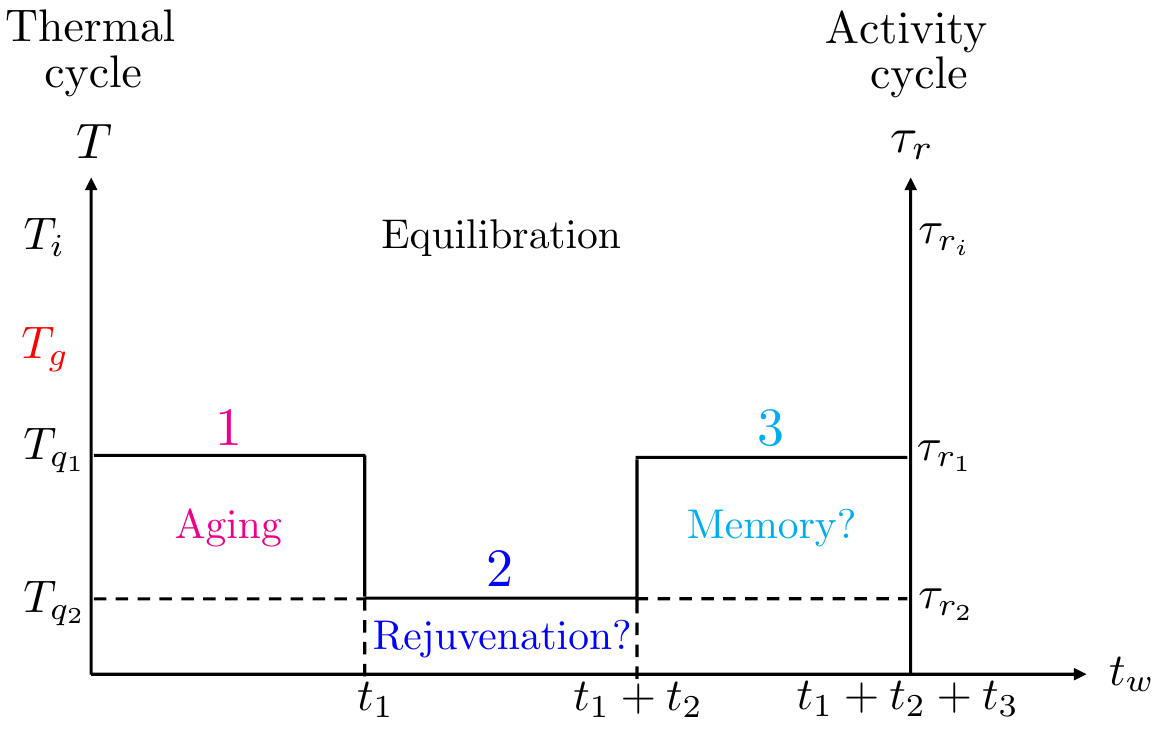} 
    \caption{Cyclic protocols used in this work to induce aging, rejuvenation, and memory in active systems. The left axis represents a thermal cycle and the right axis shows an activity cycle. }
    \label{fig:cycle}
    \end{figure}

\subsection{Protocols}
To induce possible rejuvenation and memory effects, we expose our system to two different protocols: thermal and active cycling. 
The thermal cycling protocol is applied to both an active system ($f=0.5$ and $\tau_r=1,10$) and a passive system ($f=0$) for comparison. In both cases we 
%\textbf{Thermal cycle-- } To study the effects of temperature cycling on both active systems with $f=0.5$ and $\tau_r=1,10$ and passive systems, we 
prepare $1000$ independent configurations, which we first allow to equilibrate at a high temperature $T_i=1$. We note that this relatively large number of independent trajectories is needed to ensure sufficient statistical quality of the results. Once the equilibration process is completed, we quench the temperature below the glass transition temperature to $T_{q_{1}}=0.25<T_g$, where $T_g$ is approximately $0.4$ or $0.3$ for a passive and active system, respectively \cite{flenner2015fundamental,Li_2016,janzen2021aging}. The system is then allowed to evolve at $T_{q_{1}}$ for a duration of $t_1$. %=500$ or $1000$. 
This initial aging step is followed by another quench to $T_{q_{2}}=0.1<T_{q_{2}}$, after which the system is allowed to evolve for a time $t_2$. Finally, we increase the temperature back to $T_{q_{1}}$ and allow the system to evolve for a duration of $t_3$. After some testing, we have found that $t_1=t_2=t_3=500$ or 1000 provides a reasonable time frame to probe possible rejuvenation and memory effects. The left axis of Fig.\ \ref{fig:cycle} illustrates the temperature cycle protocol used in our simulations.

For the activity cycling protocol, we %recall that the activity in our system is governed by two independent parameters, namely the self-propulsion force $f$ and the persistence time $\tau_r$. Here, we 
control the active system's behavior by quenching the persistence time $\tau_r$. To achieve this, we prepare $1000$ independent configurations at a temperature of $T=0.1$ or $0.25$ and an active force of $f=0.5$. We then equilibrate the system at a high persistence time of $\tau_{r_{i}}=100$ before quenching the persistence time to $\tau_{r_{1}}=10$. We subsequently allow the system to evolve for a time interval $t_1$. After this initial aging step, we quench the persistence time once again to $\tau_{r_{2}}=0$ to obtain a passive system. The system then evolves for a time $t_2$. Finally, we change the persistence time back to $\tau_{r_{1}}=10$ and we let the system evolve for a time interval of $t_3$ (right axis of Fig.\ \ref{fig:cycle}). In all cases we use $t_1=t_2=t_3=500$. 
We note that, as an alternative active cycling protocol, one could also quench the self-propulsion force $f$. We have verified that by using the self-propulsion force as the control parameter instead of the persistence time, similar results are found (see Supplementary Material).

To probe possible rejuvenation and memory effects, we analyze the mean-squared displacement $\langle \delta r^2(t_w,t+t_w) \rangle$ as a function of time $t$. The time $t$ ranges from 0 to the duration of the protocol's steps, which is either 500 or 1000. The waiting time $t_w$ is defined as the time elapsed after the start of the respective protocol. Explicitly, in the first step or in a direct quench, $t_w$ represents the time spent at 
the first quenching temperature $T_{q_{1}}$ or the quenched persistence time $\tau_{r_{1}}$.
%the quenching temperature or the activity used in the first step or direct quench. 
In the second and third steps of the protocol, $t_w$ is defined as $t_1+\hat{t}_w$ and $t_1+t_2+\hat{t}_w$, respectively. Here, $0 \leq \hat{t}_w < 1000$ when the duration of each step is 1000, or $0 \leq \hat{t}_w < 500$ when the duration of each step is 500.

\section{Results and Discussion}

\subsection{Rejuvenation by thermal cycling: active versus passive}
%\subsection{Thermal cycle: active versus passive}
\label{thermal-cycle}
%Here we seek to establish whether a thermal cycling can induce rejuvenation and memory effects in active and passive systems. 
%\textit{Step I: Aging ---} 
We first discuss the results of thermal cycling applied to either an active and passive thermal system.
The initial step of our thermal cycling protocol (step 1 in Fig.\ \ref{fig:cycle}) corresponds to a temperature quench from high ($T_i=1$) to low temperature ($T_{q_{1}}=0.25$), which induces simple aging. As expected, during this step, the relaxation dynamics slows down with increasing waiting time $t_w$. Specifically, both the active and passive systems exhibit a power-law scaling of the structural relaxation time with respect to the waiting time, $\tau_{\alpha} \sim t_w^{- \delta}$, where the exponent $\delta$ is different for active and passive systems. These aging results are in full agreement with literature, both for the passive and active case \cite{kob1997aging,mandal2020multiple,janzen2021aging}. 

The  
%\textbf{II step: Rejuvenation-- } Here we will focus on the 
second step of the protocol (step 2 in Fig.\ \ref{fig:cycle}) corresponds to a quench from $T_{q_{1}}=0.25$ to $T_{q_{2}}=0.1$, which may induce rejuvenation. Previous studies of passive systems \cite{PhysRevB.66.054404,PhysRevLett.122.255502} have shown that the rejuvenation effect is affected by both the temperature difference between the first and the second step $\Delta T=|T_{q_{1}}-T_{q_{2}}|$ and the duration of each step of the process. If $t_1$ is not long enough an additional aging contribution may be observed. %Therefore we will consider two different time durations, namely $t_1=500$ and $t_1=1000$. 
Testing the protocol with $t_1=t_2=500$, we have found no evidence of rejuvenation in either the passive or active system (see Supplementary Material). Indeed, during the second step of the process, with $T_{q_{2}}=0.1$, the mean-squared displacement does not evolve as the system ages. This behavior could be due to a too-short time duration or a too-small $\Delta T$. To investigate which of these parameters is responsible for this behavior, we repeat the protocol with  $t_1=t_2=1000$.

%In the passive case, even with a longer duration of each step ($t_1=1000$), the system remains frozen at the new temperature $T_{q_{2}}=0.1$, and the rejuvenation effect is not observed. The dynamics becomes slower compared to the end of the first step, and similar to the dynamics resulting from a direct quench at $T_q=T_{q_{2}}=0.1$ with a waiting time of $t_w=t_1=1000$. The aging dynamics is therefore slowed down by this additional temperature quench \cite{PhysRevResearch.3.013234}. The absence of rejuvenation could be due to a time duration that is still too short or a $\Delta T$ that is not sufficiently large. To increase $\Delta T$, we have increased $T_{q_{1}}$ while keeping $T_{q_{2}}$ constant. However, this adjustment did not alter the overall behavior, indicating that even a larger $\Delta T$ cannot induce rejuvenation in the passive system. To achieve rejuvenation, a smaller $T_{q_{2}}$ or a longer time duration $t_1$ is required \cite{PhysRevLett.122.255502}. However, decreasing the temperature or further increasing $t_1$ is computationally expensive as the dynamics become noisier, and even more ($>1000$) independent configurations would be necessary to obtain reliable results. 

As shown in the Supplementary Material, in the passive case, even with a longer duration of each step ($t_1=1000$), the system remains frozen at the new temperature $T_{q_{2}}=0.1$, and the rejuvenation effect is not observed. Different waiting times $\hat{t}_w$ exhibit the same dynamical behavior, indicating that the aging dynamics is slowed down by this additional temperature quench \cite{PhysRevResearch.3.013234}. These results are in agreement with existing literature showing that structural glasses display rejuvenation only when the duration of the first step is substantial ($t_1 = 1.2 \cdot 10^6$ or $\infty$ in Ref.\ \cite{PhysRevLett.122.255502}) and there is a considerable temperature jump between the steps. For spin glasses, especially in three dimensions, rejuvenation is also notoriously difficult to observe in simulations, and typically requires an exceptionally large temperature jump %This phenomenon is also observed in spin glasses, where an exceptionally large temperature jump was necessary for rejuvenation before the availability of Janus supercomputers 
\cite{Baity-Jesi2023}. 
Hence, in our case, we believe the absence of rejuvenation in the passive glass may be attributed to either a time duration $t_1$ that is still too short or a $\Delta T$ that is not sufficiently large. However, decreasing the temperature $T_{q_{2}}$ or further increasing $t_1$ is computationally expensive as the dynamics become noisier, and obtaining reliable results would require even more than 1000 independent configurations.
%In contrast to experiments, numerical simulations of spin-glasses have shown it is more challenging to observe rejuvenation, especially in the three-dimensional case \cite{PhysRevB.66.054404}. However, recent work by Baity-Jesi et al.\ \cite{Baity-Jesi2023} successfully replicated the phenomena of rejuvenation and memory  observed in three-dimensional experiments using more sophisticated simulations conducted on the Janus II supercomputer. The previous absence of rejuvenation in the system can be attributed to the limitations in achievable correlation lengths before the availability of Janus supercomputers. As a result, observing rejuvenation would have required an exceptionally large temperature jump. 
\begin{figure}
    \centering
    \includegraphics [width=\columnwidth] {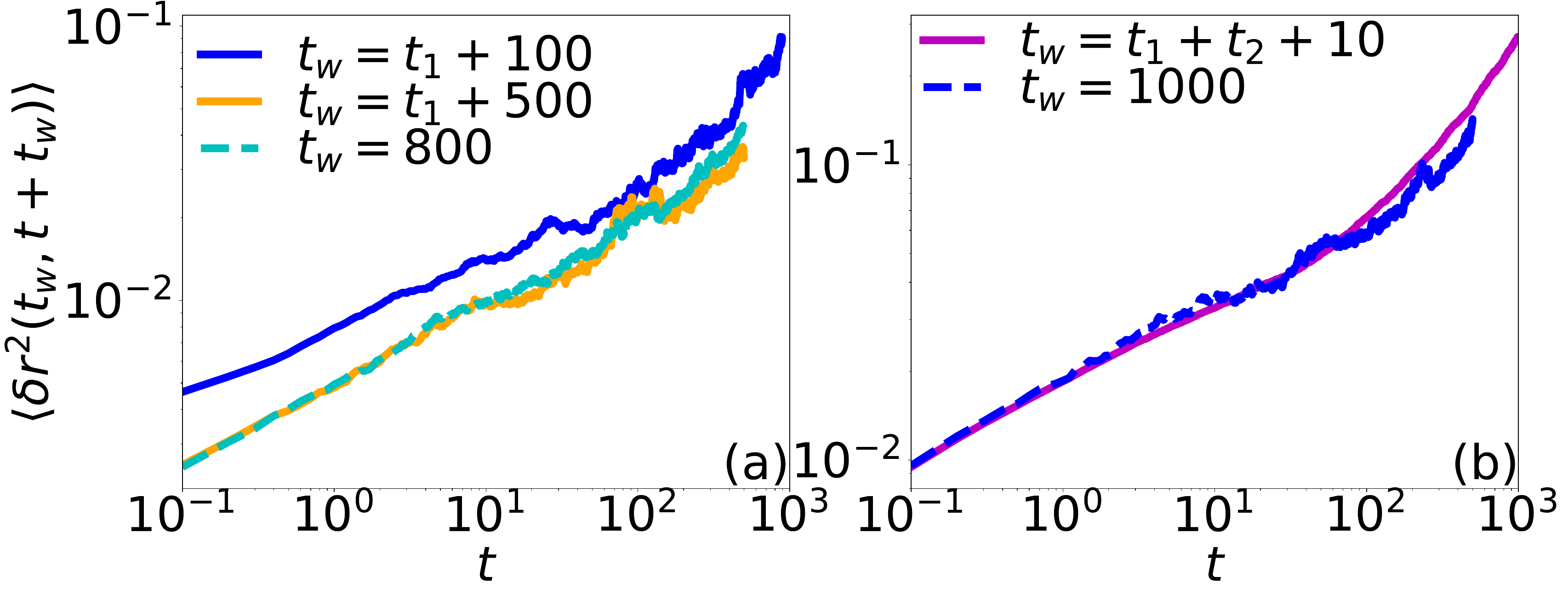} 
    \caption{Mean-squared displacements as a function of time for an active thermal system with a relatively small persistence time of $\tau_r=1$, subjected to a thermal cycle. Panels (a) and (b) correspond to the second and third step of the thermal cycling protocol to probe rejuvenation and memory effects, respectively. In both panels, solid lines correspond to different waiting times $t_w$ (with $t_1=t_2=1000$). The dashed curve in panel (a) corresponds to a direct quench to $T_{q_2}$, i.e.\ without the first aging step, at $t_w=800$; the dashed curve in panel (b) corresponds to the end of the first step, $t_w=1000$.}
    % during the second step of the thermal cycling at $T_{q_{2}}=0.1$ and $f=0.5$ for $\tau_r=1$ during  (a) the second and (b) the third step. (a) The solid curves correspond to $t_w=t_1+10,100,500$ in the second step of the cycle, i.e., the quench from $T_{q_{1}}=0.25$ to $T_{q_{2}}=0.1$. The second step takes place after a time $t_1=1000$ during which the system is at $T_{q_{1}}=0.25$. The dashed curve represents the mean-squared displacement at $t_w=800$ after a direct quench from $T_i=1$ to $T_q=0.1$. (b) The solid line represents the mean-squared displacement for $t_w=t_1+t_2+10$ in the third step of the cycle when the temperature is raised back to $T_{q_{1}}=0.25$. The dashed curve represents the mean-squared displacement at the end of the first step, $t_w=1000$.}
    \label{fig:thermal-activetau1}
    \end{figure}

Conversely, for the active system, we do observe a rejuvenation effect under the same thermal cycling protocol. To see this, let us first consider an active system with a relatively small persistence time of $\tau_r=1$.
In Fig.\ \ref{fig:thermal-activetau1}(a) we plot the mean-squared displacements for this active system during the second step of the cycle for waiting times $t_w=t_1+\hat{t}_w$, with $t_1=1000$ and $\hat{t}_w=100,500$. For comparison we also show the dynamics following a direct quench to the same temperature, $T_q=T_{q_{2}}=0.1$.
%To establish whether rejuvenation takes place for an active system with $\tau_r=1$ and $f=0.5$, we compute the mean-squared displacement for $t_w=t_1+10,100,500$. In Fig. \ref{fig:thermal-activetau1}(a) we compare the behavior during the second step of the cycle with a direct quench at $T_q=T_{q_{2}}=0.1$. 
Notably, we observe that the mean-squared displacement at $t_w=t_1+500$ overlaps almost perfectly with that of a directly-quenched sample at $t_w=800$. This is the hallmark of rejuvenation:
even though the system during the second step of the cycle is older (age $t_w=1500$), it behaves effectively as a younger system (age $t_w=800$) that was quenched to the same temperature. 
%Overall, the behavior obtained from a thermal cycle is thus equivalent to a direct quench with an effective waiting time 
%$t_w^{eff}$ , which in this case is equal to 800. This effective waiting time can be defined as 
%$t_w^{eff}=\hat{t}_w+t^{ag}$, where $t^{ag}$ is nonzero but smaller than $t_1$ (in this case $t_w^{eff}=800$ and $t^{ag}=300$). Therefore, $t^{ag}$ is the only consequence of the time spent at $T_{q_{1}}$.  Note that the limiting case of  $t^{ag}=0$ would indicate perfect rejuvenation, i.e., as if the first aging step never happened.  
Since the behavior obtained from a thermal cycle at waiting time $t_1+\hat{t}_w$ is equivalent to a direct quench at waiting time $t_w$, this implies that the effective age of the system during the second step is $t_w^{eff}=t_w$. Consequently, in this case, the effective waiting time can be expressed as $t_w^{eff}=\hat{t}_w+t^{ag}$, where $t^{ag}$ is nonzero but smaller than $t_1$ (in this case $t_w^{eff}=800$ and $t^{ag}=300$).
Note that the limiting case of  $t^{ag}=0$ would indicate perfect rejuvenation, i.e., as if the first aging step never happened. 
%Overall, the behavior obtained from a thermal cycle is thus equivalent to a direct quench with an effective waiting time $t_w^{eff}=t_w+t^{ag}$, where in this case $t_w$ is the waiting time corresponding to a direct quench and $t^{ag}$ is nonzero but smaller than $t_1+t_w$ (in this case $t^{ag}=300$). Note that the limiting case of $t^{ag}=0$ would indicate perfect rejuvenation, i.e., as if the first aging step never happened.  

\begin{figure}
    \centering
    \includegraphics [width=\columnwidth] {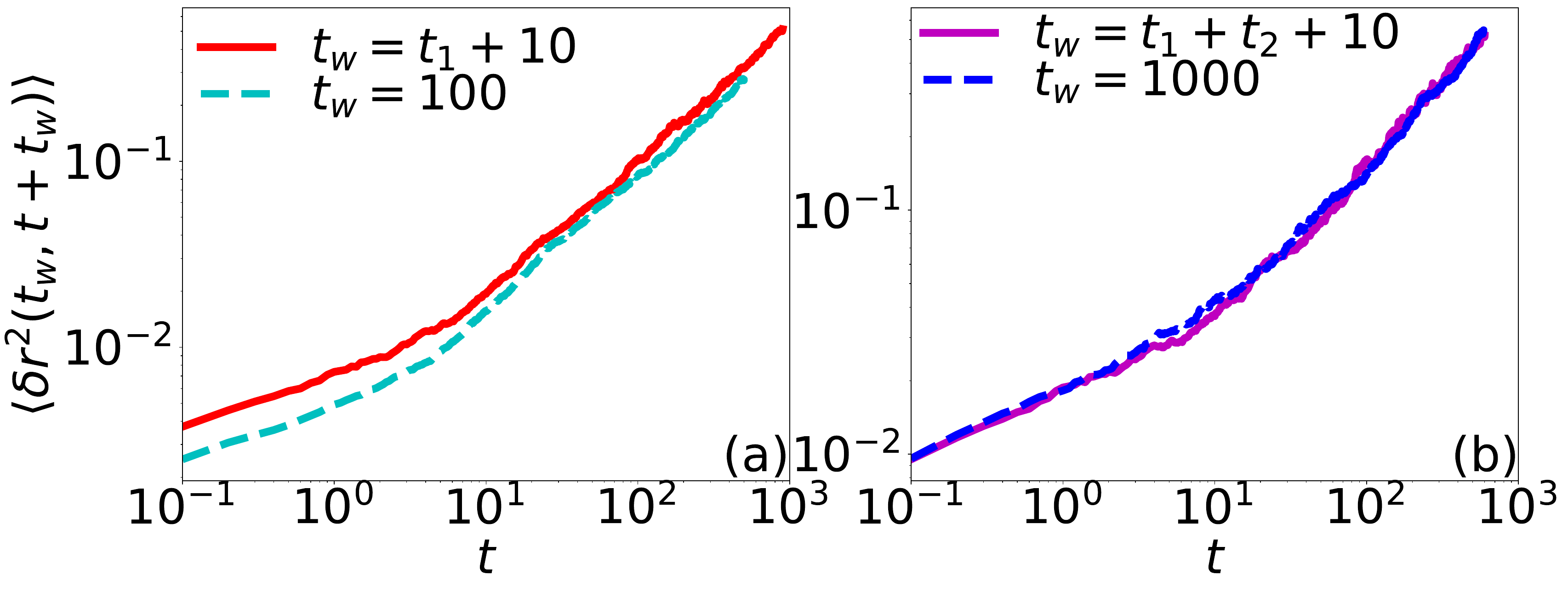} 
    \caption{Mean-squared displacements as a function of time for an active thermal system with a relatively large persistence time of $\tau_r=10$, subjected to a thermal cycle. Panels (a) and (b) correspond to the second and third step of the thermal cycling protocol to probe rejuvenation and memory effects, respectively. In both panels, solid lines correspond to different waiting times $t_w$ (with $t_1=t_2=1000$). The dashed curve in panel (a) corresponds to a direct quench to $T_{q_2}$, i.e.\ without the first aging step, at $t_w=100$; the dashed curve in panel (b) corresponds to the end of the first step, $t_w=1000$.}
    %Mean-squared displacement as a function of time during the second step of the thermal cycling at $T_{q_{2}}=0.1$ and $f=0.5$ for $\tau_r=10$ during  (a) the second and (b) the third step. (a) The solid curves correspond to $t_w=t_1+10,100,500$ in the second step of the cycle, i.e., the quench from $T_{q_{1}}=0.25$ to $T_{q_{2}}=0.1$. The second step takes place after a time $t_1=1000$ during which the system is at $T_{q_{1}}=0.25$. The dashed curve represents the mean-squared displacement at $t_w=800$ after a direct quench from $T_i=1$ to $T_q=0.1$. (b) The solid line represents the mean-squared displacement for $t_w=t_1+t_2+10$ in the third step of the cycle when the temperature is raised back to $T_{q_{1}}=0.25$. The dashed curve represents the mean-squared displacement at the end of the first step, $t_w=1000$.}
    \label{fig:thermal-activetau10}
    \end{figure}

%indicates that, e%ven though the system during the second step of the cycle is effectively older than a direct quench at the same temperature, . Thus, the behavior obtained from a thermal cycle is equivalent to a direct quench with an effective waiting time $t_w^{eff}=t_w+t^{ag}$, where $t^{ag}$ is nonzero but smaller than $t_1+t_w$. %This is the hallmark of rejuvenation, since 
To further investigate the rejuvenation effect in active systems, we consider a larger persistence time of $\tau_r=10$.
Figure \ref{fig:thermal-activetau10}(a) shows the corresponding mean-squared displacements at a waiting time $t_w=t_1+10$ during the second step of the cycle, and at a waiting time $t_w=100$ following a direct quench to the same temperature $T_{q_{2}}$. When comparing the two curves, %behavior during the second step with a direct quench at the same temperature, 
we find that the short-time behavior of the mean-squared displacements at $t_w=t_1+10$ and $t_w=100$ differs, in particular, the system in the second step is initially faster than a direct quench. Consequently, the short-time dynamics is still influenced by the first step of the protocol. However, the long-time behaviors of the mean-squared displacements at $t_w=t_1+10$ and $t_w=100$ overlap. Therefore, we can conclude that increasing the persistence time in active systems leads to a 'stronger' rejuvenation effect. In this case, the extra aging contribution $t^{ag}$ remains present, but is smaller compared to the one observed for the system with the smaller persistence time of $\tau_r=1$. As a result, the system with $\tau_r=10$ is effectively younger than the system with $\tau_r=1$.

%To further investigate the rejuvenation effect in active systems, we consider a larger persistence time of $\tau_r=10$.
%Figure \ref{fig:thermal-activetau10}(a) shows the corresponding mean-squared displacements at waiting times $t_w=t_1+10$ (second step of the cycle). When comparing the behavior during the second step with a direct quench at the same temperature, we find that the long-time behavior of the mean-squared displacements at $t_w=t_1+10$ and $t_w=100$ overlap. In this case, the extra aging contribution $t^{ag}$ is still present but smaller compared to the one found for the smaller persistence time, implying that the system with $\tau_r=10$ is effectively younger than the system with $\tau_r=1$. Therefore we can conclude that increasing the persistence time in active systems leads to a 'stronger' rejuvenation effect. 

Overall our results of the second step show that an active glassy system exhibits rejuvenation under thermal cycling, while its passive counterpart exposed to the same protocol does not. However, it must be noted that the rejuvenation process is 'weak' in the sense that the material is still effectively older than a system subjected to a direct quench. In other words, the effect of the first aging step is still noticeable in the dynamics. %Moreover, the persistence time of the active particles plays a crucial role in the strength of the rejuvenation effect. 
We hypothesize that the active rejuvenation effect is due to the ability of self-propelling particles to escape their cages relatively easily, thus allowing them to effectively restart aging at the new temperature. To rationalize the importance of the persistence time on the rejuvenation dynamics, we note that the dynamics of a thermal active glass is governed by both thermal and active effects. On sufficiently long timescales, however,  it is known that the activity becomes dominant \cite{janzen2021aging}, and hence systems with larger persistence times exhibit stronger rejuvenation effects.

\subsection{Memory by thermal cycling:  active versus passive}
%\textbf{III step: Memory-- } 
Let us now look at the emergence of memory effects due to a temperature cycle. Such memory effects can occur in the final step of the thermal cycling protocol (step 3 in Fig. \ref{fig:cycle}), in which the temperature is raised from $T_{q_{2}}=0.1$ to $T_{q_{1}}=0.25$. Figures \ref{fig:thermal-activetau1}(b) and \ref{fig:thermal-activetau10}(b) show the mean-squared displacements for two active systems with  $\tau_r=1$ and $10$, respectively, at $t_w=t_1+t_2+10$ (third step of the cycle) and $t_w=1000$ (at the end of the first step). In both cases, the results reveal that the system recovers the same behavior observed at the end of the first step after a waiting time $\hat{t}_w=10$. Thus, for the waiting time $t_w^{mem}=t_1+t_2+t^{*}$, the system retains a memory of the time spent at $T_{q_{1}}$. However, due to the relatively small value of $t^{*}=10$ (for both $\tau_r=1$ and $10$), the memory is not perfect, and this is the only consequence of the time spent at $T_{q_{2}}$. This behavior is consistent with previous results obtained in (passive)spin glasses \cite{vincent1995contrasting,Vincent2007}, where it was shown that the time needed to remember the time spent at $T_{q_{1}}$ is much shorter compared to the time spent at $T_{q_{2}}$ and hence, after a very brief transient, the dynamics continues as if the second step has never occurred. Moreover, our results indicate that, unlike the rejuvenation effect, the memory effect is independent of the persistence time. In order to achieve full memory, i.e.\ $t^{*}=0$, we hypothesize that the temperature jump $\Delta T$ should  be increased, as it has been shown in spin-glasses that increasing $\Delta T$ leads to a decrease in $t^{*}$ \cite{vincent1995contrasting,Vincent2007}.

As shown in the Supplementary Material, the memory effect can also be observed in passive systems, even when rejuvenation is absent. 
%even though the rejuvenation effect is not observed when the duration of each step is equal to 500 for both active and passive systems, or in the passive case when $t_1=1000$, the memory effect is present. 
During the first step, the dynamics is frozen at $T_{q_{2}}$, but shortly after the temperature is raised back to $T_{q_{1}}$, the passive system remembers its behavior during the first step. In a similar manner to the active case, the memory effect in passive systems also exhibits a small $t^{*}$, which represents the only consequence of the second step of the cycling. This suggests that, unlike rejuvenation, the memory effect is easier to observe in these systems.
The presence of memory effects in these systems, even without rejuvenation, is consistent with the findings in the existing literature, where memory represents a common phenomenon observed in various systems. Such systems include materials experiencing cyclic deformation and biologically relevant systems like blood flow and growing tissue monolayers \cite{SpecialTopicMemoryformation}.

%\subsection{Active cycle: from active to passive}
\subsection{Rejuvenation and memory by active cycling: from active to passive}

We now consider the activity cycling protocol that, in contrast to thermal cycling, can only be applied to active systems. Here it is instructive to note that the cyclic change of the ABP persistence time $\tau_r$ also corresponds to an \textit{effective} temperature cycle \cite{bechinger2016active}.%, since the effective temperature of an ABP is related to $\tau_r$ via 
%$T_{eff}=T+\frac{\gamma f^2 \tau_r}{2k_B}$ \cite{bechinger2016active}. 
Hence, this protocol offers not only a unique means to generate novel dynamics in active systems, but it also allows us to induce effectively large temperature jumps between steps via $\tau_r$ (while keeping $T$ fixed). The latter aspect is particularly useful to induce rejuvenation on reasonable simulation time scales, as rejuvenation demands a large temperature difference between steps or long step durations (see Sec.\ \ref{thermal-cycle}). 

%As discussed in the previous section, the rejuvenation effect is only observed when there is a large temperature difference between steps or long step durations. However, decreasing the temperature increases noise, and longer steps increase computational costs. To avoid increasing noise in the data, we can perform an activity cycling making the system passive and then active again. By doing so we quench the effective temperatures \cite{bechinger2016active}, allowing for a larger temperature jump between steps.

%In the dilute case, the active system can be mapped into a  passive system at an effective temperature defined as $T_{eff}=T+\frac{\gamma f^2 \tau_r}{2k_B}$, \cite{bechinger2016active}. By increasing the active force $f$ or the persistence time $\tau_r$, the effective temperature can be increased. When the active force is zero or the persistence time is zero, the passive case is recovered. To avoid increasing noise in the data, we can quench the effective temperatures, allowing for a larger temperature jump between steps. Therefore, we apply a persistence time cycle, which involves equilibrating the system at a high persistence time of $\tau_r=100$. At time zero, the persistence time is quenched to $\tau_{r_{1}}=10$, and after a time $t_1=500$, the persistence time is quenched again to $\tau_{r_{2}}=0$. Finally, after $t_2=t_1$, the persistence time is raised again to $\tau_{r_{1}}=10$.

     \begin{figure}
    \centering
    \includegraphics [width=\columnwidth] {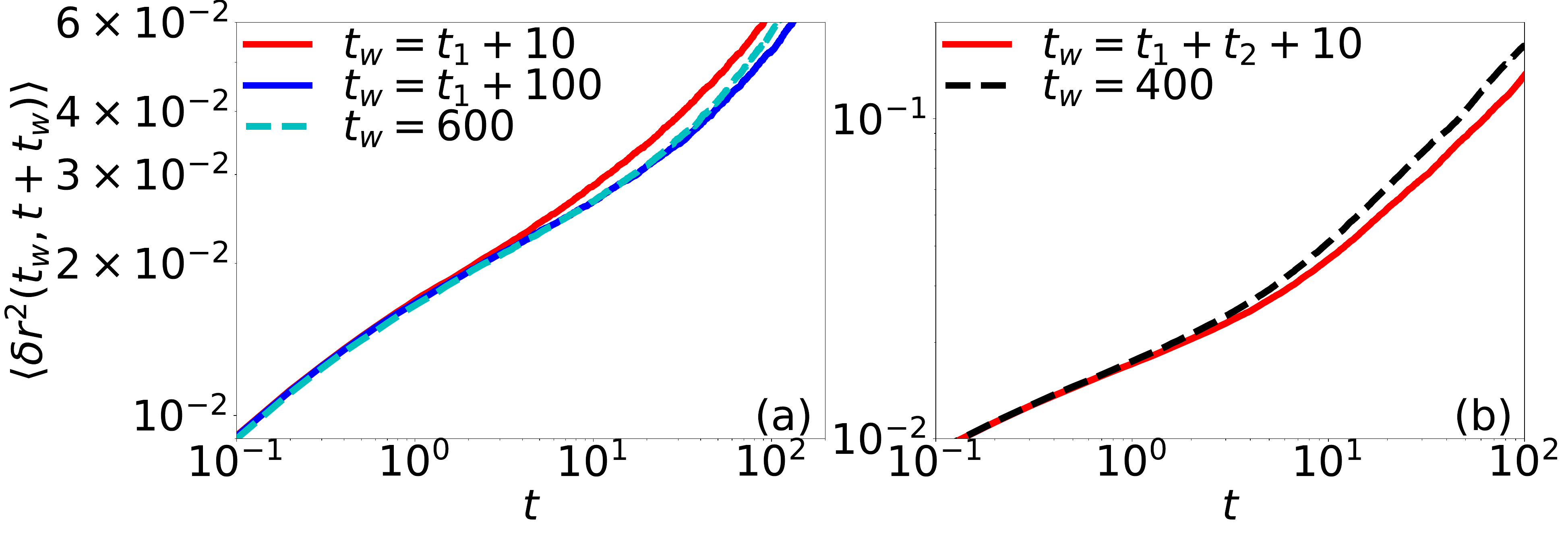} 
    \caption{Mean-squared displacements as a function of time for an active thermal system ($T=0.25, f=0.5$) subjected to an activity cycle. Panels (a) and (b) correspond to the second and third step of the active cycling protocol to probe rejuvenation and memory effects, respectively. In both panels, solid lines correspond to different waiting times $t_w$ (with $t_1=t_2=500$). 
    The dashed curve in panel (a) corresponds to a direct quench to a passive system at $t_w=600$; the dashed curve in panel (b) represents the mean-squared displacement before the end of the first step  $t_w=400$.}
    \label{fig:taur-cycle-T025}
    \end{figure}
    
During the first step of the activity cycle, we again observe simple aging, similar to the thermal cycling scenario. Figure~\ref{fig:taur-cycle-T025}(a) illustrates the mean-squared displacement in the second step of the activity cycling (i.e.\ where the persistence time is quenched to $\tau_{r_2}=0$) at a fixed temperature of $T=0.25$. Unlike thermal cycling, where the dynamics is frozen at $t_1=500$, in activity cycling, the dynamics evolves for different $\hat{t}_w$. Notably, the aging restarts, and at a waiting time of approximately $t_w = t_1 + 100$, the system recovers the behavior observed after a direct quench at $\tau_{r_{2}}=0$. The behavior in the second step overlaps with the one found with a direct quench for $t_w=600$, thus indicating a rejuvenation effect. Here, since $t_1=500$ and $\hat{t}_w=100$, we can conclude that applying a direct quench or an activity cycling is equivalent, but the first step of the cycling has an impact on the second step because $t^{ag}=t_1$. We have verified that the same rejuvenation behavior can be found when using the self-propulsion force as the control parameter instead of the persistence time (see Supplementary Material). This indicates that this behavior is protocol-independent, and quenching the self-propulsion force or the persistence time leads to the same dynamical behavior.
    
\begin{figure}
    \centering
    \includegraphics [width=\columnwidth] {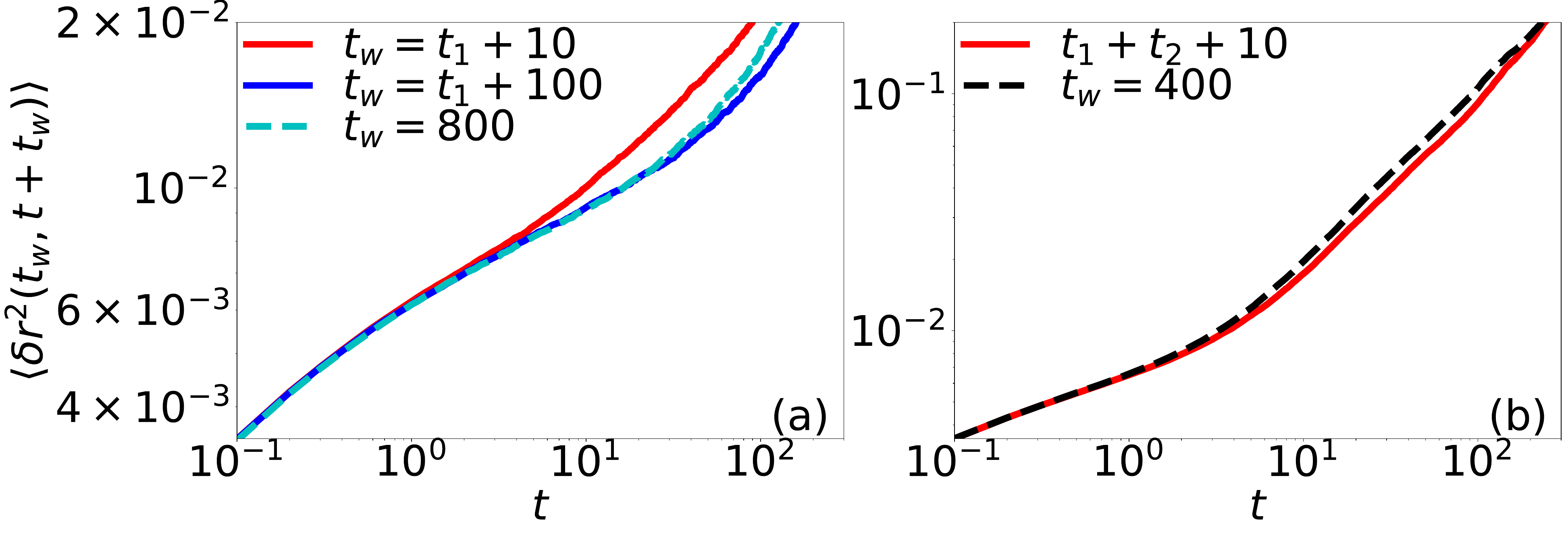} 
    \caption{Mean-squared displacements as a function of time for an active thermal system ($T=0.1, f=0.5$) subjected to an activity cycle. Panels (a) and (b) correspond to the second and third step of the active cycling protocol to probe rejuvenation and memory effects, respectively. In both panels, solid lines correspond to different waiting times $t_w$ (with $t_1=t_2=500$). 
    The dashed curve in panel (a) corresponds to a direct quench to a passive system at $t_w=800$; the dashed curve in panel (b) represents the mean-squared displacement before the end of the first step $t_w=400$.}
   
    \label{fig:taur-cycle-T01}
    \end{figure}

To understand if temperature plays a role in this weak rejuvenation effect, we repeat the activity cycling at a lower temperature of $T=0.1$. Figure~\ref{fig:taur-cycle-T01}(a) shows the behavior of the system during the second step of the activity cycling at this lower temperature. We find that the behavior of the system at the second step at $t_w=t_1+100$ overlaps with that observed from a direct quench at $t_w=800$.
This indicates that the consequence of the first step of the cycle makes the system slightly older compared to a direct quench. This phenomenon, known as \textit{overaging} in the literature \cite{PhysRevLett.89.065701,Physics_beyond_Aging}, can be attributed to the fact that $t_1$ is not long enough.
Comparing this activity cycling to thermal cycling, we observe that the short duration of the first step leads to overaging and freezing of the dynamics, respectively. In conclusion, the observation of non-equilibrium phenomena such as rejuvenation becomes more accessible when activity cycling is applied. This is attributed to the significantly larger effective temperature jump compared to a purely passive system undergoing thermal cycling.

Finally, as shown in  Fig.~\ref{fig:taur-cycle-T025}(b) and  Fig.~\ref{fig:taur-cycle-T01}(b), in the third step, when the persistence time is raised again to $\tau_{r_{1}}=10$, the system quickly recovers the same behavior found at the end of the first step. Consequently, the system has a memory of the first step as if the second step did not happen. In agreement with the thermal cycle, the system needs a time $t^*$ to 'remember' the time spent at $\tau_{r_{1}}$. The presence of this $t^*$  can be due to the fact that the duration of each step of the cycle is not long enough. Therefore, we conclude that when applying an active cycle, we observe memory effects. We have also verified that this memory effect is present when the control parameter of the activity cycle is the self-propulsion force (see Supplementary Material). In contrast to the rejuvenation effect, the memory effect found within the activity cycle is consistent with the one observed when thermal cycling is applied.

\section{Conclusions} 
 In summary, our work reveals that the non-equilibrium dynamic behaviors caused by a temperature cycle in an active thermal system are significantly different from those in a passive system. Specifically, an active system subjected to a temperature cycle exhibits rejuvenation and memory effects, whereas in the passive case, the system gets frozen at the new temperature, and the rejuvenation effect is absent \cite{PhysRevResearch.3.013234}. Nevertheless, even in the absence of rejuvenation in the passive case, a memory effect similar to that observed in the active system can still be found. Additionally, we find that the rejuvenation effect becomes stronger as the persistence time increases, whereas the memory effect is independent of this parameter. We can rationalize the enhanced rejuvenation of more persistent active particles by considering that the long-time behavior of active thermal glasses is dominated by activity rather than thermal motion \cite{janzen2021aging}. Hence, a more active system with a larger persistence time will more easily restart the aging process after a sudden temperature change, leading to a 'stronger' rejuvenation effect. %The presence of such out-of-equilibrium dynamics in an active glass can be attributed to the activity-enhanced cage breaking \cite{ni2013pushing} that ensures a restart of the aging, leading to rejuvenation. 

We find that rejuvenation and memory effects in active matter can also be induced by a non-equilibrium protocol unique to active systems, namely via an activity cycle. In particular, the application of an activity cycle from an active to a passive system enables access to higher temperature jumps without increasing the noise, leading to rejuvenation. Depending on the temperature of the system, we can observe either rejuvenation or overaging, with the latter occurring at lower temperatures. To mitigate the overaging effect at low temperatures, it is necessary to increase the duration of the first step of the protocol. Overall, our findings suggest that activity cycling offers richer non-equilibrium dynamics in the second step of the cycle compared to thermal cycling, when the duration of the first step is short. Moreover, even with this protocol, the memory of the first step of the cycle is quickly recovered when the system becomes active again. Our results also demonstrate that similar outcomes can be achieved by employing the self-propulsion force as the control parameter in the activity cycle. Moreover, these findings align with prior research indicating that different protocols yield comparable dynamics. This similarity is reminiscent of quenching protocols employed to study simple aging, where an activity quench generates analogous outcomes to a temperature quench \cite{janzen2021aging,mandal2020multiple}.

Our study provides new insights into the rich non-equilibrium dynamics of active glasses and emphasizes the significance of exploring the interplay between temperature and activity. Furthermore, our results show that, although the nature of an active system differs from that of spin glasses, the out-of-equilibrium dynamics observed during a temperature cycle exhibit remarkable similarities. The rejuvenation and memory effects observed in spin glasses have been attributed to the slow increase of a characteristic length following a temperature quench \cite{PhysRevB.38.373,koper1988domain,PhysRevLett.58.57}, or explained using a hierarchical energy-landscape picture \cite{mezard1987spin,Parisi1980,PhysRevLett.35.1792,PhysRevLett.63.2853,Jin2017}. However, since active systems are non-Hamiltonian and are not governed by energy minimization, understanding the presence of rejuvenation and memory effects in these systems will require a different theoretical framework. Finally, given that dense active matter is becoming increasingly relevant in the context of biology, it will be interesting to explore in future work how physical aging, rejuvenation, and memory are manifested in biological glassy systems such as confluent cell layers, tissues, and solid tumors \cite{janssen2019active,angelini2011glass,garcia2015physics,schoetz2013glassy,atia2021cell,D1SM01291F,PhysRevX.13.031003}. 

\section{Acknowledgments} 
We thank Kees Storm for his critical reading of the manuscript. This work has been financially supported by the Dutch Research Council (NWO) through a Physics Projectruimte grant.
\bibliographystyle{apsrev4-1} 
\bibliography{./aging}
\end{document}